\documentclass[aps,pra,preprint,showpacs,floatfix,endfloats*]{revtex4}
\usepackage{amsmath}

\begin{document}

\title{Excitation energies, hyperfine constants, E1, E2, M1 transition rates,
and  lifetimes of $6s^2nl$ states  in Tl~I and Pb~II }

\author{ U. I. Safronova}
\email{usafrono@nd.edu}
 \affiliation{Physics Department, University of Nevada, Reno, NV
 89557}

\author{M. S. Safronova}
\email{msafrono@udel.edu} \affiliation {Department of Physics and
Astronomy, 223 Sharp Lab,
 University of Delaware, Newark, Delaware 19716}

\author{W. R. Johnson}
\email{johnson@nd.edu} \homepage{www.nd.edu/~johnson}
\affiliation{Department of Physics,  University of Notre Dame,
Notre Dame, IN 46556}

\date{\today}
\begin{abstract}
Energies of $6s^2np_j$ ($n$ = 6--9),  $6s^2ns_{1/2}$ ($n$ = 7--9),
 $6s^2nd_j$
($n$ = 6--8), and $6s^2nf_{5/2}$ ($n$ = 5--6) states in Tl~I and
Pb~II are obtained using relativistic many-body perturbation
theory. 
  Reduced matrix elements,
oscillator strengths, transition rates, and lifetimes are
determined for the 72 possible $6s^2nl_j-6s^2n'l'_{j'}$
electric-dipole transitions. Electric-quadrupole and
magnetic-dipole matrix elements are evaluated to obtain
$6s^2np_{3/2} - 6s^2mp_{1/2}$ ($n,m=6,7$) transition rates.
Hyperfine  constants $A$  are evaluated for $6s^2np_j$
($n$ = 6--9), $6s^2ns_{1/2}$ ($n$ = 7--9), and
 $6s^2nd_j$ ($n$ = 6--8)
  states  in $^{205}$Tl.
First-, second-, third-, and
all-order corrections to the energies and matrix elements 
and first- and second-order Breit corrections to energies
are calculated.
In our implementation of the all-order method, 
 single and double
excitations of Dirac-Fock wave functions
are included to all orders in perturbation theory. 
These calculations provide a theoretical benchmark 
for comparison with experiment and theory.
 \pacs{31.15.Ar, 31.15.Md, 32.10.Fn, 32.70.Cs}
\end{abstract}
\maketitle
\section{Introduction}

The ground-state energy of thallium, treated as a
one-electron system, was calculated  by \citet{dzuba88-tl}
using perturbation theory in a screened Coulomb interaction (PTSCI), by
\citet{tl-en} using  third-order many-body
perturbation theory (MBPT), and by \citet{kelly-tl} using
the coupled-cluster (CC) approach. 
Second-order MBPT energies of thallium were
evaluated for the ground and  excited states ($6p_{1/2}$,
$6p_{3/2}$, $7s_{1/2}$, and  $7p_{1/2}$) in \citet{tl-hyp-90},
where hyperfine constants for the four states listed above and 
electric-dipole transition matrix elements $7s-6p_{1/2}$,
$7s-7p_{1/2}$, and $7s-6p_{3/2}$ were also evaluated.  In all of the above
calculations the three-electron state $6s^2nl$  with 78
core electrons was considered as a one-electron
$nl$ system with an 80 electron core [Hg].

Thallium was treated as three-particle system in \citet{3elec},
where energies of low-lying excited states were calculated using
second-order relativistic MBPT starting from a frozen core $V^{(N-3)}$
HF potential.
A combined MBPT plus
configuration-interaction (CI) method was employed in
\citet{dzuba87-tl,dzuba-ci} to evaluate  the Tl~I ionization
potential and the first few energy levels. This  method was also
used recently by \citet{kozlov} to calculate  energies and
hyperfine constants for seven low-lying states, dipole
matrix elements between those states, and the parity 
nonconserving (PNC)  E1 amplitude for the
$6p_{1/2}-6p_{3/2}$ transition in $^{205}$Tl. A review of all
previous papers regarding calculations of the PNC E1
amplitude in Tl is given in Ref.~\cite{kozlov} and therefore omitted from the
present discussion.

The lifetime of the $7\, ^2S_{1/2}$ state in Tl was measured
using the optical double-resonance technique and
the level-crossing technique by \citet{tl-64}; this lifetime was also
measured
using the zero-field level-crossing (Hanle effect) technique by \citet{tl-71}.
Lifetimes of the $n\, ^2P_j$ states in Tl~I $(n = 7-11)$ were
determined in Refs.~\cite{tl-7p85,tl-6p86} through measurements of
the fluorescence signal resulting from two-photon excitation in an
atomic beam. Level-crossing spectroscopy with pulsed two-photon
excitation was used in \cite{tl-hyp-88,tl-pra88,tl-hypd} to
determine lifetimes for  $n\, ^2P_{3/2}$ $(n=8,9)$ states
 and hyperfine splitting for $n\, ^2P_{3/2}$ $(n=8-11)$ states. 
 Measurement of the electric-quadrupole amplitude
for the 1283~nm $6\, ^2P_{1/2}-6\, ^2P_{3/2}$ transition in Tl~I
was reported by \citet{tl-6p99}. Spectroscopic studies of the
plasma generated in a thallium arc were reported by Alonso-Medina
in Refs.~\cite{tl-96,tl-97}. Relative transition probabilities for
the  18 infrared lines arising from excited doublets in Tl~I
($n\, ^2S_{1/2} - 7\, ^2P_j$, $n\, ^2D_{3/2}-7\, ^2P_j$, 
$n\, ^2P_{3/2}-6\, ^2D_j$) were determined from emission line-intensity
measurements of an optically-thin light source \cite{tl-96}. The
optical emission spectra (2000 \AA-15000 \AA) of a plasma produced in
a Tl arc lamp were recorded and analyzed to
obtain relative transition probabilities for 26 lines ($n\,
^2S_{1/2} - 6\, ^2P_J$, $n\, ^2D_{J'}-6\,^2P_J$, $n\, ^2P_{J}-7\,
^2S_{1/2}$) in Tl~I  \cite{tl-97}.

Lifetimes of the $7\, ^2S_{1/2}$,  $7\, ^2P_{J}$,  $7\,
^2D_{3/2}$, $5\, ^2F_{J}$, and $6\,^2F_{5/2}$ levels of Pb~II were
measured by \citet{pb-life} using the multichannel method of
delayed coincidences. The lifetime of the metastable $6\,
^2P_{3/2}$ level of Pb~II was measured experimentally by
\citet{pb-6p} using the ion storage technique and time-resolved
registration of the subsequent decay photons. Relative transition
probabilities of 30 Pb~II lines arising from excited doublet
levels were obtained by Alonso-Medina in Refs.~\cite{pb-96,pb-97}.
Transition probabilities  for 30 lines involving $S$, $P$, $D$,
and $F$ levels of Pb~II were determined from a hollow cathode
discharge by \citet{pb-96}. In Refs.~\cite{pb-97,pb-01a},
transition probabilities of Pb~II were determined from the values
of intensities obtained using the emission of a plasma
generated by focusing a laser beam on a Pb target. Calculations of
Pb~II properties including a potential model to represent the core
polarization of Pb~II  were reported by \citet{pb-migdalek} and by
\citet{pb-96}. Oscillator strengths and relative line strengths
for the $6\, ^2P_{J} - n\, ^2S_{1/2}$ transitions  $(n = 7-11)$
and $7\, ^2P_{J} - n\, ^2S_{1/2}$ transitions with $(n = 8-11)$ were
presented in Ref.~\cite{pb-migdalek}. Transition probabilities for
the  $7\, ^2P_{J} - n\, ^2S_{1/2}$  $(n = 8-11)$, $7\, ^2P_{J} - n\,
^2D_{J'}$  $(n = 7, 8, 10)$, $7\, ^2S_{1/2} - 7\, ^2P_{J}$, and $6\,
^2D_{J} - n\, ^2F_{J'}$  $(n = 5-7)$ transitions were calculated
by \citet{pb-96}. The relativistic Hartree-Fock (HFR) Cowan code
was used by \citet{forb-96}, to calculate the $6\, ^2P_{1/2} - 6\,
^2P_{3/2}$ magnetic-dipole and electric-quadrupole transition
probabilities in Tl~I and Pb~II. Magnetic-dipole and
electric-quadrupole transition rates for  $n\, ^2P_{1/2} - n'\,
^2P_{3/2}$ transitions $(n,n' = 6, 7)$ in Tl were presented
by \citet{tl-forb-77}, where valence-electron wave functions were
generated as numerical solutions of the Dirac equation in a
modified Tietz central potential. Transition probabilities for 190
lines arising from the $6s^2ns\, ^2S_{1/2}$, $6s^2np\, ^2P_{J}$,
$6s^2nd\, ^2D_{J}$, $6s^2nf\, ^2F_{J}$, $6s6p^2\, ^{2}S_{1/2},\
^{2,4}P_{J},\ ^{2,4}D_{J}$ levels in Pb~II were calculated
recently by \citet{pb-01b} also using  the Cowan code.
Theoretical lifetimes  were also calculated in \cite{pb-01b}
 for those levels for which lifetimes measurements were
given in \cite{pb-life}. 
Pb~II was detected with the Goddard High Resolution Spectrograph
aboard the Hubble Space Telescope in several stars \cite{star} and
 radiative properties
of Pb~II  are needed to derive various astrophysical parameters.

In the present paper, energies of the $6s^2np_j$ $(n = 6-9)$,
$6s^2ns_{1/2}$ $(n = 7-9)$, $6s^2nd_j$ $(n = 6-8)$, and
$6s^2nf_{5/2}$ $(n = 5-6)$ states in  Tl~I and Pb~II are obtained
using relativistic MBPT. We carry out third-order and all-order 
calculations of energies, transition amplitudes, and hyperfine constants.
Second-order Breit corrections to energies
are also calculated.  Here, we treat the three-electron system $6s^2nl$ 
with a 78 electron core as a 
one-electron system $nl$ with an 80 electron [Hg] core,
permitting us to use
experience from previous MBPT studies of atomic properties in
systems with one valence electron
Refs.~\cite{li-en,na-en,cu-en,agmar,blundell-li,Liu,
blundell-cs,safr-na,safr-alk,gold}.
  Reduced matrix elements,
oscillator strengths, transition rates, and lifetimes are
determined for the 72 possible $nl_j-n'l'_{j'}$ electric-dipole
transitions. Electric-quadrupole and magnetic-dipole matrix
elements are evaluated for $np_{1/2}-mp_{3/2}$ $(m,n=6,7)$ transitions.
 We investigate the hyperfine structure in
 $^{205}$Tl to determine the hyperfine constants $A$ for the $6s^2np_j$
$(n = 6-9)$, $6s^2ns_{1/2}$ $(n= 7-9)$, and
 $6s^2nd_j$ $(n= 6-8)$ states in $^{205}$Tl.

\section{Third-order MBPT calculations of energies of Tl~I and Pb~II}

Results of our third-order calculations of energies, which are
carried out following the method described in \cite{agmar,gold},
are summarized in Table~\ref{tab1}, where we list lowest-order
Dirac-Fock energies $E^{(0)}$, first-order Breit energies
$B^{(1)}$, second-order Coulomb $E^{(2)}$ and Breit $B^{(2)}$
energies, third-order Coulomb energies  $E^{(3)}$, single-particle
Lamb shift corrections $E_\text{ LS}$, and the sum of the above
$E_{\text{tot}}$. The first-order Breit energies include
retardation, whereas the second-order Breit energies are evaluated
using the unretarded Breit operator. The Lamb shift is
approximated as the sum of the one-electron self energy and the
first-order vacuum-polarization energy. The vacuum-polarization
contribution is calculated from the Uehling potential using the
results of Fullerton and Rinker \cite{vacuum}. The self-energy
contribution is estimated for the $s$, $p_{1/2}$ and $p_{3/2}$
orbitals by interpolating among the values obtained by
\citet{mohr1,mohr2,mohr3} using Coulomb wave functions. For this
purpose, an effective nuclear charge $Z_\text{eff}$ is obtained by
finding the value of $Z_\text{eff}$ required to give a Coulomb
orbital with the same average $\langle r\rangle$ as the Dirac-Hartree-Fock
(DHF) orbital.

We find that the correlation corrections to energies in neutral Tl
and
 Tl-like Pb are large, especially for the $6p$ states. For
example, $E^{(2)}$ is about 20\% of $E^{(0)}$ and $E^{(3)}$ is about
40\% of $E^{(2)}$ for the $6p_{j}$ states of neutral Tl. 
Despite the evident slow
convergence of the perturbation theory expansion, the $6p_{1/2}$
energy from the third-order MBPT calculation is within
0.2\% of the measured ionization energy for the $6p_{1/2}$ state
of neutral Tl and improves for higher valence states and for
Pb$^+$ (0.03\%). The order of levels changes from Tl to Pb$^+$.
For example, the $5f_{5/2}$ and $6f_{5/2}$ states, which are in
the thirteenth and nineteenth places for neutral Tl, are in the
ninth  and fifteenth places for Pb$^+$. In two last columns of
Table~\ref{tab1}, we compare our results for the energy levels of
the 19  single-particle states of interest in Tl~I and Pb~II with
recommended values from the National Institute of Standards and
Technology (NIST) database \cite{nist}. Although our results are
generally in good agreement with the NIST data, discrepancies are
found. One cause for these discrepancies is the omission of the
fourth- and higher-order correlation corrections in
the theoretical values. A second possible
cause is the omission of three-particle interactions in our
single-particle model space. The importance of $6s^26d$ + $6s6p^2$
mixing for Tl~I was emphasized by \citet{3elec}. Moreover,
\citet{pb-01b} included the $6s6p^2$ states in calculations used
to classify the observed spectrum of Pb~II.

We use B-splines \cite{Bspline} to generate a complete set of
basis DHF wave functions for use in the evaluation of MBPT
expressions. For  Tl~I and Pb~II, we use 40 splines of order $k=7$
for each angular momentum. The basis orbitals are constrained to
cavities of radii $R=85$ a.u.\ and $R=65$ a.u.\ for  Tl~I and
Pb~II, respectively. The cavity radius is chosen large enough to
accommodate all $6l_j$ and $5f_j$ orbitals considered here and
small enough that 40 splines can approximate inner-shell DHF wave
functions with good precision. We include 35 out of 40 basis orbitals
for each partial wave in our third-order energy calculations,
since the contributions from the five highest-energy orbitals are
negligible. The second-order calculation includes partial waves up
to $l_{\text{max}}=8$ and is extrapolated to account for
contributions from higher partial waves.  A lower number of
partial waves, $l_{\text{max}}=6$, is used in the third-order
calculation.  Since the asymptotic $l$-dependence of the second-
and third-order energies are similar (both fall off as $l^{-4}$),
we use the second-order remainder as a guide when extrapolating
the third-order energy.

\section{All-order SD calculations of Tl~I and Pb~II energies}

Results of our all-order SD calculations for the 19 lowest states
in neutral Tl and Tl-like Pb ion are presented in Table~\ref{tab-esd}, 
where we
list lowest-order (DHF) energies $E^{(0)}$, SD correlation
energies $E^\text{{SD}}$, omitted third-order terms
$E^{(3)}_\text{{extra}}$,
  first- and second-order Breit corrections to energies $B^{(n)}$, $n$ = 1, 2, single-particle
Lamb shift corrections $E_\text{ LS}$,
 totals $E_\text{ tot}$, and values from NIST, $E_\text{{NIST}}$
\cite{nist}. The method used to evaluate $E^\text{{SD}}$ 
is described in 
Refs.~\cite{blundell-li,Liu,blundell-cs,safr-na,safr-alk,gold}.
The SD equations are set up in a finite basis and solved
iteratively to give the single- and double-excitation coefficients
and the correlation energy $E^\text{{SD}}$. Contributions
$E^{(3)}_\text{{extra}}$
 in Table~\ref{tab-esd} account for that part of the
third-order MBPT correction not included in the SD energy
(see Eq.~(2.16) in Ref.~\cite{Safronova}). The basis orbitals used
to define single-particle  states are linear combinations of
B-splines as in the third-order calculation above.

Comparing  Tables~\ref{tab1} and \ref{tab-esd} shows that the
second-order term $E^{(2)}$ is the dominant contribution to
$E^\text{{SD}}$, as expected.  We have already mentioned the
importance of higher partial wave contributions to $E^{(2)}$ and
$E^{(3)}$. In the SD calculation, we include partial waves through
$l$=6 in $E^\text{{SD}}$ and use the difference 
$E^{(2)}$ - $E^{(2)}_{l\leq 6}$ to extrapolate our result.

The column labeled $\delta E$ in Table~\ref{tab-esd} gives
differences between our {\it ab initio}
 results and the recommended values \cite{nist}.  
The SD results agree better with the recommended
values than do the third-order MBPT results, except
for the ionization potential. This confirms our previous comments
concerning the slow convergence of the perturbation theory and
illustrates the importance of fourth- and higher-order correlation
corrections. Those differences between the present theoretical
results and the recommended values that were not
improved by the SD method are most probably due to the omission of
three-particle interactions mentioned previously.

\section{Electric-dipole matrix elements, oscillator strengths, transition
rates, and lifetimes in Tl~I and Pb~II}

Transition matrix elements provide another test of the quality of
atomic-structure calculations and another measure of the size of
correlation corrections.  Reduced electric-dipole matrix
elements between low-lying states of Tl~I and Pb~II calculated to
third order and in the SD approximation
are presented in
Table~\ref{tab-dip}.  First-order reduced matrix elements
$Z^{(1)}$ are obtained from  DHF calculations,
second-order reduced matrix elements  $Z^{(2)}$
 include $Z^{(1)}$ and the second-order
correction associated with the random-phase approximation (RPA).
All results given Table~\ref{tab-dip} are obtained in length form
for matrix elements. Length-form and velocity-form matrix elements
differ typically by 1 - 10\% for DHF matrix elements and 1 - 3 \%
for the second-order matrix elements.

 The third-order matrix elements $Z^{(3)}$ include
$Z^\text{ (RPA)}$ (all higher-order RPA corrections), third-order
Bruekner-orbital $Z^\text{ (BO)}$, structural radiation $Z^\text{
(SR)}$, and normalization $Z^\text{ (NORM)}$ corrections,
described in Refs.~\cite{dip3,Safronova}.  We find
second-order  RPA corrections to be  very large, 10-40\%, being
the smallest for the $7p_{j}$-$7d_{j'}$ transitions.

Electric-dipole matrix elements evaluated in the SD approximation
are given in  columns labeled  $Z^\text{{(SD)}}$  in
Table~\ref{tab-dip}. A detailed discussion of the calculations of
matrix elements in the SD approximation is found in
Refs.~\cite{blundell-li,safr-na,safr-alk,Safronova,gold}. It
should be noted that SD matrix elements $Z^\text{{SD}}$ include
$Z^{(3)}$ completely, along with important fourth- and
higher-order corrections. The fourth-order corrections omitted
from SD matrix elements were discussed recently by \citet{der-4}.

In Table~\ref{tab-dip-com}, we compare our SD data
$Z^{\text{(SD)}}$  for reduced matrix elements of the dipole
operator in Tl~I  with theoretical $Z^{\text{(theor)}}$
 and experimental $Z^{\text{(expt)}}$  data
 given by \citet{kozlov} and references therein.
  Our
 SD data are in excellent agreement with experimental and
 theoretical data  from Refs.~\cite{dzuba-ci,kozlov}
  obtained by combining MBPT and CI methods.

Transition rates $A_r$ (s$^{-1}$), oscillator strengths ($f$),
and line strengths $S$ (a.u.) for transitions in Tl~I and Pb~II
calculated in SD approximation  are summarized in
Tables~\ref{tab-osc-tl} and \ref{tab-osc-pb}, respectively.  It
should be noted that we use theoretical energies obtained in the SD
approximation in the evaluation of transition rates and oscillator
strengths. For convenience, we present wavelengths for all
transitions in Tables~\ref{tab-osc-tl} and 
\ref{tab-osc-pb} calculated using the SD approximation.  
The largest oscillator
strengths agree with experimental results within the corresponding
uncertainties in many cases. 

In Table~\ref{tab-ar-com} we compare transition rates $A_r$
(10$^6$s$^{-1}$) and wavelengths  $\lambda$ (\AA) in Tl~I and
Pb~II with available experimental measurements given in
Refs.~\cite{tl-64,tl-96,tl-97} for Tl~I and in
Refs.~\cite{pb-01b,pb-01a} for Pb~II. The SD dipole matrix
elements and energies are used to obtain results given in columns
 labeled  $A_{r}^{\text{(SD)}}$ and $\lambda^{\text{(SD)}}$.
Our SD results for the wavelengths are in excellent agreement with
experimental wavelengths; the discrepancies  are  about
0.01--0.1\%. Experimental results for transition rates were given
in Refs.~\cite{tl-64,tl-96,tl-97,pb-01b,pb-01a} with an estimated error 
of 10\%. Our SD values for transition rates agree in many cases with
experimental results within the experimental uncertainties.  
However, we found huge disagreements  for two transitions,
$7s_{1/2}-8p_{1/2}$ and $6d_{3/2}-8p_{1/2}$ in Pb~II. For the
first  transition, $7s_{1/2}-8p_{1/2}$,  the RPA contribution is
very important (compare $Z^{(1)}$ and $Z^{(2)}$ in
Table~\ref{tab-dip}); for the second  transition,
$6d_{3/2}-8p_{1/2}$, the BO contribution is largest
($Z^{(1)}$=0.3065, $Z^\text{(BO)}$=-0.3063)
cancelling almost exactly the lowest-order contribution. Such
large correlation contributions are responsible for the large difference
between our the third-order and SD results and the lowest-order
DHF results. All previous calculations for transitions rates in
Pb~II were based on the relativistic Hartree-Fock (HFR) Cowan
code (Refs.~\cite{pb-96,pb-01b}) and those data were used for
experimental estimates of transition rates presented in column
with heading $A_{r}^{\text{(expt.)}}$ of Table~\ref{tab-ar-com}.

We calculate lifetimes of the $np_j$, $ns_{1/2}$ $(n = 7-9)$,
 $nd_j$ $(n = 6-8)$, and $nf_{5/2}$ $(n = 5-6)$ states in  Tl~I and Pb~II
 using both third-order MBPT and SD results
for dipole matrix elements and energies. We list lifetimes
$\tau^\text{(SD)}$ obtained using SD method in
Table~\ref{tab-life}. In this table,
 we compare our calculated lifetimes with
available experimental measurements that are primarily obtained
for $np_j$ levels. The experimental data   for Tl~I are from
Refs.~\cite{tl-71,tl-6p86}  and the Pb~II data  are from
Ref.~\cite{pb-01b} and references therein. Experimental results
for $\tau^\text{expt}$ were given in Refs.~\cite{tl-6p86} and
~\cite{pb-01b} with 5\% and 10\% errors, respectively.  Our SD
results are in excellent agreement  with experimental results
   except for one
case, $7s_{1/2}$ state in Pb~II. This is unexpected  since we have
perfect agreement between the $\tau^\text{SD}$ and
$\tau^\text{expt}$ from Ref.~\cite{tl-71} for
$7s_{1/2}$ state in Tl~I. As one can  see from
Table~\ref{tab-life},  the values of lifetime in Tl~I are
 larger than the values of lifetime in Pb~II
by a factor of 4. In this case, it seems strange that
$\tau^\text{expt}$ for the $7s_{1/2}$ state in Pb~II is almost
identical with $\tau^\text{expt}$ for the $7s_{1/2}$ state in Tl~I.

\section{Electric-quadrupole and magnetic-dipole  transitions
 in Tl~I and Pb~II}

Reduced matrix elements of the electric-quadrupole (E2) and
magnetic-dipole (M1) operators in lowest, second, third, and all
orders of perturbation theory in  Tl~I and Pb~II are given in
Table~\ref{tab-dip-e2m1}. Detailed descriptions of the
calculations  of the  reduced matrix elements of the
E2 and M1 operators in lowest and
second orders of  perturbation theory were given by \citet{ca}.
Third-order and all-order calculations are done in the same
way as the calculations of E1 matrix elements. In
Table~\ref{tab-dip-e2m1}, we present  E2 and M1 matrix elements
in the $Z^{(1)}$, $Z^{(2)}$, $Z^{(3)}$, and $Z^\text{(SD)}$
approximations  for $6p_{1/2}-6p_{3/2}$, $6p_{1/2}-7p_{3/2}$,
$7p_{1/2}-6p_{3/2}$, and $7p_{1/2}-7p_{3/2}$ transitions in Tl~I
and Pb~II. These four transitions in Tl~I were investigated for
the first time by \citet{tl-forb-77}. The importance of the Breit
contribution to the calculation of  the $np_{j}-n'p_{j'}$ matrix
elements in Tl~I was underlined in Ref.~\cite{tl-forb-77}.
 We
found also that  the  second-order Breit contribution is larger
than the Coulomb contribution, which are unusually small
 for above mentioned E2 and M1 transitions. 
As a result, the difference between $Z^{(1)}$ and
$Z^{(2)}$ presented in Table~\ref{tab-dip-e2m1} is rather small;
about 0.3--1\%.   The largest Coulomb contribution arises from
Bruekner-orbital $Z^\text{(BO)}$ correction which
is especially large for E2 matrix elements in Tl; the
ratio $Z^\text{ (BO)}$/$Z^{(1)}$ is about 0.13--0.25.

  Wavelengths $\lambda$ (\AA),
transition rates for E2 transitions $A^{E2}_{r}$ and
M1 transitions  $A^{M1}_{r}$
 (s$^{-1}$) in Tl~I and Pb~II calculated in
the SD approximation are presented in Table~\ref{tab-ar-e2m1}. The SD
data ($a$) are compared with theoretical calculations
  given in Refs.~\cite{tl-forb-77,forb-96}. The differences between
  our results and those from \cite{tl-forb-77,forb-96} for $A^{E2}_{r}$ and
   $A^{M1}_{r}$ can be explained by the additional correlation correction taken
   into account in the SD approximation. 

\section{Hyperfine constants for neutral thallium}

Calculations of hyperfine constants follow the same pattern as 
calculations of reduced dipole matrix elements described in the
previous section. The magnetic moment and nuclear spin used in
the present calculations are taken from \cite{web}.  In
Tables~\ref{contr-hyp} and \ref{tab-hyp}, we give the
magnetic-dipole hyperfine constants $A$ for $^{205}$Tl
 and compare them with available theoretical and experimental data from
Refs.~\cite{kozlov,tl-forb-77,tl-pra88} and references therein.

Contributions to SD values for the $6p_j$, $6d_j$, and $7s_{1/2}$
states in  $^{205}$Tl are given in Table~\ref{contr-hyp} using
designations from Refs.~\cite{blundell-li,Safronova}. The first
and last lines of  Table~\ref{contr-hyp} give Dirac-Fock and SD
values of hyperfine constants $A^{(\rm HF)}$ and $A^{(\rm SD)}$,
respectively. Values of individual contributions to the 
SD matrix elements, $Z^{(a)}$--$Z^{(t)}$ in the notation of 
Ref.~\cite{Safronova}, are also given in the table.
As one can see from Table~\ref{contr-hyp}, the
largest contribution  for $6p_j$
and $7s_{1/2}$ states are from $Z^{(a)}$ and $Z^{(c)}$. The
term $Z^{(a)}$ contains second- and third-order RPA
contributions and $Z^{(c)}$ contains the part of the third-order
Bruekner orbital correction. The three first terms $Z^{(a)}$,
$Z^{(b)}$, and $Z^{(c)}$ give  more than 90\% of the  correlation
contribution for $6p_{1/2}$ and $7s_{1/2}$ states.

In Table~\ref{tab-hyp}, we list hyperfine constants
 $A$ for $^{205}$Tl
 and compare our values with available
theoretical and experimental data from
Refs.~\cite{kozlov,tl-forb-77,tl-pra88}
 and references therein. In this table, we present
 the first-order $A^\text{(DHF)}$
and all-order $A^\text{(SD)}$  values for the $np_j$ with $n$ =
6--9, $ns_{1/2}$ with $n$ = 7--9, and $6d_j$ levels. The largest
disagreements between our SD data and experimental values occur for
$6p_{3/2}$ and $8d_{5/2}$ states. As we mentioned previously, the
largest contribution for the $6p_{3/2}$ state comes from the
second- and third-order RPA contribution  and this
$Z^{(a)}$ term cancels the lowest-order $A^{(\rm DHF)}$ term. 
 With such cancellation it is difficult to calculate
$A(6p_{3/2}$) accurately.
 The best agreement with experimental measurements is found for
$6p_{1/2}$ and $7p_{1/2}$ states, 0.4\% and 0.3\%, respectively.
For other states ($7s_{1/2}$, $7p_{3/2}$, $6d_{j}$, $8p_{j}$ and
$9p_{j}$) the discrepancies  range from 2 to 7\%.

\section{Conclusion}
In summary,  a systematic relativistic  MBPT study of the energies
 of $6s^2np_j$ $(n = 6-9)$,  $6s^2ns_{1/2}$ $(n= 7-9)$,
 $6s^2nd_j$
$(n= 6-8)$, and $6s^2nf_{5/2}$ $(n= 5-6)$ states in  Tl~I and
Pb~II  is presented. The energy values are in good agreement with
existing experimental energy data and provide a theoretical
reference database for the line identification. A systematic
 all-order SD study of reduced matrix elements,
oscillator strengths, and transition rates for the 72 possible
$6s^2nl_j-6s^2n'l'_{j'}$ electric-dipole transitions is conducted.
 Electric-quadrupole and magnetic-dipole
matrix elements are evaluated to calculate lifetime of
$6s^26p_{3/2}$ state. Hyperfine constants are presented for
$6s^2np_j$ $(n = 6-9)$, $6s^2ns_{1/2}$ $(n = 7-8)$, and
 $6s^2nd_j$ $(n = 6-8)$
  states  in $^{205}$Tl isotope.
We believe that our energies and transition rates will be useful
in analyzing existing experimental data and in planning future
measurements.

\begin{acknowledgments}
The work of W.R.J.  was supported in part by National Science
Foundation Grant No.\ PHY-01-39928.  The work of U.I.S. was
supported by DOE/NNSA under UNR grant DE-FC52-01NV14050.
\end{acknowledgments}

\newpage

\begin{table*}
\caption{\label{tab1} Zeroth-order (DHF), second-, and third-order
Coulomb correlation energies $E^{(n)}$,
  first- and second-order Breit corrections $B^{(n)}$, and  Lamb shift $E_\text{
  LS}$ contributions to the energies of Tl~I and Pb~II.
The total energies $E_\text{ tot}$ for Tl~I and  Pb~II are
compared with experimental
 energies $E_\text{{NIST}}$ \protect\cite{nist}, ($\delta E$ = $E_\text{
 tot}$ - $E_\text{{NIST}}$). Units: cm$^{-1}$.}
\begin{ruledtabular}
\begin{tabular}{lrrrrrrrrr}
\multicolumn{1}{c}{$nlj$ } &
\multicolumn{1}{c}{$E^{(0)}$} &
\multicolumn{1}{c}{$E^{(2)}$} &
\multicolumn{1}{c}{$E^{(3)}$} &
\multicolumn{1}{c}{$B^{(1)}$} &
\multicolumn{1}{c}{$B^{(2)}$} &
\multicolumn{1}{c}{$E_\text{ LS}$}&
\multicolumn{1}{c}{$E_\text{ tot}$} &
\multicolumn{1}{c}{$E_\text{{NIST}}$} &
\multicolumn{1}{c}{$\delta E$} \\
\hline
\multicolumn{9}{c}{Tl~I}\\
$6p_{1/2}$&  -43824& -7741&  2534&  259&  -421&  -2&  -49194& -49264&     70\\
$6p_{3/2}$&  -36636& -6739&  2414&  135&  -291&   0&  -41117& -41471&    354\\
$7s_{1/2}$&  -21109& -2022&   663&   26&   -46&   3&  -22484& -22787&    303\\
$7p_{1/2}$&  -14276& -1033&   306&   29&   -42&   0&  -15017& -15104&     87\\
$7p_{3/2}$&  -13357&  -965&   316&   19&   -38&   0&  -14025& -14103&     78\\
$6d_{3/2}$&  -12218&  -959&   296&    3&    -8&   0&  -12886& -13146&    260\\
$6d_{5/2}$&  -12167&  -921&   280&    3&    -8&   0&  -12813& -13064&    251\\
$8s_{1/2}$&  -10040&  -602&   205&    9&   -15&   0&  -10444& -10518&     74\\
$8p_{1/2}$&   -7599&  -370&   102&   11&   -16&   0&   -7872&  -7896&     24\\
$8p_{3/2}$&   -7249&  -352&   106&    7&   -14&   0&   -7502&  -7523&     21\\
$7d_{3/2}$&   -6864&  -404&   133&    2&    -4&   0&   -7138&  -7253&    115\\
$7d_{5/2}$&   -6837&  -387&   120&    1&    -4&   0&   -7107&  -7224&    117\\
$5f_{5/2}$&   -6863&   -98&    25&    0&     0&   0&   -6936&  -6948&     12\\
$9s_{1/2}$&   -5893&  -264&    91&    4&    -7&   0&   -6069&  -6098&     29\\
$9p_{1/2}$&   -4741&  -179&    49&    5&    -8&   0&   -4873&  -4884&     11\\
$9p_{3/2}$&   -4569&  -172&    50&    4&    -7&   0&   -4694&  -4702&      8\\
$8d_{3/2}$&   -4391&  -216&    71&    1&    -2&   0&   -4537&  -4591&     54\\
$8d_{5/2}$&   -4376&  -198&    61&    1&    -2&   0&   -4514&  -4572&     58\\
$6f_{5/2}$&   -4393&   -56&    15&    0&     0&   0&   -4434&  -4441&      7\\
\multicolumn{9}{c}{Pb~II}\\
$6p_{1/2}$& -114546& -9441&  2947&  460&  -623&  -3& -121206&-121243&     37\\
$6p_{3/2}$& -100787& -8782&  3007&  264&  -472&   1& -106769&-107162&    393\\
$7s_{1/2}$&  -58728& -3901&  1331&   71&  -108&  10&  -61325& -61795&    470\\
$6d_{3/2}$&  -46790& -4454&  1354&   35&   -74&   0&  -49929& -51503&   1574\\
$6d_{5/2}$&  -46259& -4257&  1262&   28&   -73&   0&  -49300& -52279&   2979\\
$7p_{1/2}$&  -44847& -2369&   694&   79&   -94&   0&  -46539& -46784&    245\\
$7p_{3/2}$&  -42277& -2067&   614&   52&   -84&   0&  -43761& -43971&    210\\
$8s_{1/2}$&  -30980& -1370&   473&   28&   -42&   2&  -31889& -32063&    174\\
$5f_{5/2}$&  -27730& -1101&   340&    0&    -4&   0&  -28495& -28729&    234\\
$7d_{3/2}$&  -26282& -1657&   401&   16&   -33&   0&  -27555& -26959&   -596\\
$7d_{5/2}$&  -26040& -1610&   368&   13&   -33&   0&  -27302& -25939&  -1363\\
$8p_{1/2}$&  -25378&  -963&   286&   33&   -39&   0&  -26061& -26236&    175\\
$8p_{3/2}$&  -24296&  -856&   252&   22&   -35&   0&  -24912& -25005&     93\\
$9s_{1/2}$&  -19246&  -649&   221&   14&   -20&   1&  -19680& -19897&    217\\
$6f_{5/2}$&  -17796&  -639&   203&    0&    -3&   0&  -18235& -18375&    140\\
$8d_{3/2}$&  -16895&  -827&   152&    8&   -17&   0&  -17578& -17849&    271\\
$8d_{5/2}$&  -16767&  -810&   134&    7&   -17&   0&  -17453& -17053&   -400\\
$9p_{1/2}$&  -16412&  -497&   148&   17&   -20&   0&  -16765& -16821&     56\\
$9p_{3/2}$&  -15850&  -453&   131&   12&   -20&   0&  -16181& -16223&     42\\
\end{tabular}
\end{ruledtabular}
\end{table*}

\begin{table*}
\caption{\label{tab-esd} Lowest-order (DHF) $E^{(0)}$,
single-double Coulomb  $E^\text{{SD}}$, $E^{(3)}_\text{{extra}}$,
  first- and second-order Breit corrections $B^{(n)}$, and Lamb shift $E_\text{
  LS}$ contributions to the energies of Tl~I and Pb~II.
The total energies $E_\text{ tot}$ for Tl~I and  Pb~II are
compared with experimental
 energies $E_\text{{NIST}}$ \protect\cite{nist}, ($\delta E$ = $E_\text{
 tot}$ - $E_\text{{NIST}}$). Units: cm$^{-1}$.}
\begin{ruledtabular}
\begin{tabular}{lrrrrrrrrr}
\multicolumn{1}{c}{$nlj$ } &
\multicolumn{1}{c}{$E^{(0)}$} &
\multicolumn{1}{c}{$E^\text{{SD}}$} &
\multicolumn{1}{c}{$E^{(3)}_\text{{extra}}$} &
\multicolumn{1}{c}{$B^{(1)}$} &
\multicolumn{1}{c}{$B^{(2)}$} &
\multicolumn{1}{c}{$E_\text{ LS}$}&
\multicolumn{1}{c}{$E_\text{ tot}$} &
\multicolumn{1}{c}{$E_\text{{NIST}}$} &
\multicolumn{1}{c}{$\delta E$} \\
\hline
\multicolumn{9}{c}{Tl~I}\\
  $6p_{1/2}$& -43824& -5767& 694&259&-421& -2&-49061&  -49264& 203\\
  $6p_{3/2}$& -36636& -5301& 807&135&-291&  0&-41286&  -41471& 185\\
  $7s_{1/2}$& -21109& -1939& 266& 26& -46&  3&-22799&  -22787& -12\\
  $7p_{1/2}$& -14276&  -904& 103& 29& -42&  0&-15091&  -15104&  13\\
  $7p_{3/2}$& -13357&  -768& 109& 19& -38&  0&-14034&  -14103&  69\\
  $6d_{3/2}$& -12218& -1098& 152&  3&  -8&  0&-13169&  -13146& -23\\
  $6d_{5/2}$& -12167& -1062& 147&  3&  -8&  0&-13088&  -13064& -24\\
  $8s_{1/2}$& -10040&  -576&  79&  9& -15&  1&-10542&  -10518& -24\\
  $8p_{1/2}$&  -7599&  -308&  34& 11& -16&  0& -7877&   -7896&  19\\
  $8p_{3/2}$&  -7249&  -280&  37&  7& -14&  0& -7499&   -7523&  24\\
  $7d_{3/2}$&  -6864&  -447&  64&  2&  -4&  0& -7250&   -7253&   3\\
  $7d_{5/2}$&  -6837&  -434&  62&  1&  -4&  0& -7212&   -7224&  12\\
  $5f_{5/2}$&  -6863&   -97&  16&  0&   0&  0& -6945&   -6948&   3\\
  $9s_{1/2}$&  -5893&  -234&  34&  4&  -7&  0& -6096&   -6098&   2\\
  $9p_{1/2}$&  -4741&  -147&  17&  5&  -8&  0& -4874&   -4883&   9\\
  $9p_{3/2}$&  -4569&  -136&  18&  4&  -8&  0& -4691&   -4702&  11\\
  $8d_{3/2}$&  -4391&  -227&  33&  1&  -2&  0& -4586&   -4591&  5\\
  $8d_{5/2}$&  -4376&  -221&  32&  1&  -2&  0& -4567&   -4572&  5\\
  $6f_{5/2}$&  -4393&   -55&   9&  0&   0&  0& -4439&   -4441&   2\\
\multicolumn{9}{c}{Pb~II}\\
  $6p_{1/2}$&  -114546&-6958&  732&  460&-623&  -3&-120937& -121243& 306\\
  $6p_{3/2}$&  -100787&-6740&  934&  264&-472&   1&-106800& -107162& 362\\
  $7s_{1/2}$&   -58728&-3726&  510&   71&-108&  10& -61970&  -61795&-175\\
  $6d_{3/2}$&   -46790&-5086&  664&   35& -74&   0& -51251&  -51503& 252\\
  $6d_{5/2}$&   -46259&-5051&  647&   28& -73&   0& -50708&  -52279&1571\\
  $7p_{1/2}$&   -44847&-2072&  254&   79& -94&   0& -46681&  -46784& 103\\
  $7p_{3/2}$&   -42277&-1787&  227&   52& -84&   0& -43868&  -43971& 103\\
  $8s_{1/2}$&   -30980&-1202&  181&   28& -42&   2& -32012&  -32063&  51\\
  $5f_{5/2}$&   -27730&-1070&  170&    0&  -4&   0& -28634&  -28729&  95\\
  $7d_{3/2}$&   -26282&-1413&  241&   16& -33&   0& -27471&  -26959&-512\\
  $7d_{5/2}$&   -26040&-1396&  241&   13& -33&   0& -27215&  -25939&-1276\\
  $8p_{1/2}$&   -25378& -833&  106&   33& -39&   0& -26111&  -26236& 125\\
  $8p_{3/2}$&   -24296& -740&   95&   22& -35&   0& -24954&  -25005&  51\\
  $9s_{1/2}$&   -19246& -592&   86&   14& -20&   1& -19757&  -19897& 140\\
  $6f_{5/2}$&   -17796& -605&   98&    0&  -3&   0& -18307&  -18375&  68\\
  $8d_{3/2}$&   -16895& -759&  119&    8& -17&   0& -17543&  -17849& 306\\
  $8d_{5/2}$&   -16767& -761&  121&    7& -17&   0& -17417&  -17053&-364\\
  $9p_{1/2}$&   -16412& -429&   56&   17& -20&   0& -16788&  -16821&  33\\
  $9p_{3/2}$&   -15850& -387&   50&   12& -20&   0& -16196&  -16223&  27\\
\end{tabular}
\end{ruledtabular}
\end{table*}

\begin{table*}
\caption{\label{tab-dip} Reduced electric-dipole matrix elements
 in first, second, third, and all orders of
perturbation theory in  Tl~I and Pb~II.}
\begin{ruledtabular}
\begin{tabular}{llrrrrrrrrr}
\multicolumn{2}{c}{Transition}&
\multicolumn{1}{c}{$Z^{(1)}$ }&
\multicolumn{1}{c}{$Z^{(2)}$ }&
\multicolumn{1}{c}{$Z^{(3)}$ }&
\multicolumn{1}{c}{$Z^\text{{(SD)}}$ }&
\multicolumn{1}{c}{}&
\multicolumn{1}{c}{$Z^{(1)}$ }&
\multicolumn{1}{c}{$Z^{(2)}$ }&
\multicolumn{1}{c}{$Z^{(3)}$ }&
\multicolumn{1}{c}{$Z^\text{{(SD)}}$ }\\
\hline
\multicolumn{5}{c}{Tl~I}&
\multicolumn{5}{c}{Pb~II}\\
  $6p_{1/2}$&$  7s_{1/2}$&   2.0484&    1.9122&    1.6613&    1.8200&&   1.3753&    1.2966&    1.1715&    1.0085\\
  $6p_{3/2}$&$  7s_{1/2}$&   3.9655&    3.6551&    3.0780&    3.3945&&   2.6208&    2.3611&    2.2085&    2.0529\\
  $7s_{1/2}$&$  7p_{1/2}$&   6.6178&    6.3827&    5.8869&    5.9038&&   4.8098&    4.4599&    4.2312&    3.9493\\
  $7s_{1/2}$&$  7p_{3/2}$&   8.7938&    8.5490&    7.9345&    7.8706&&   6.5091&    6.1060&    5.7792&    5.3768\\
  $6p_{1/2}$&$  6d_{3/2}$&   2.7215&    2.4497&    2.3033&    2.3739&&   2.8176&    2.3669&    2.1956&    2.0661\\
  $6p_{3/2}$&$  6d_{3/2}$&   1.6334&    1.4983&    1.3628&    1.4193&&   1.5267&    1.3053&    1.2081&    1.2177\\
  $7p_{1/2}$&$  6d_{3/2}$&  11.9837&   11.7386&   10.5970&   10.5775&&   6.3430&    6.0271&    5.2770&    5.3111\\
  $7p_{3/2}$&$  6d_{3/2}$&   5.3951&    5.3130&    4.7860&    4.7470&&   2.7044&    2.6096&    2.2487&    2.2590\\
  $6p_{3/2}$&$  6d_{5/2}$&   4.8402&    4.4464&    3.8568&    4.1692&&   4.5018&    3.8655&    3.6018&    3.4345\\
  $7p_{3/2}$&$  6d_{5/2}$&  16.2960&   16.0456&   14.5260&   14.4021&&   8.3319&    8.0293&    6.9817&    6.9668\\
  $6p_{1/2}$&$  8s_{1/2}$&   0.6431&    0.5767&    0.5593&    0.5312&&   0.4587&    0.4170&    0.3415&    0.3709\\
  $6p_{3/2}$&$  8s_{1/2}$&   0.9779&    0.8286&    0.7392&    0.7736&&   0.7357&    0.5929&    0.5375&    0.5414\\
  $7p_{1/2}$&$  8s_{1/2}$&   6.4883&    6.4998&    6.1030&    6.2513&&   3.5698&    3.6096&    3.4487&    3.4602\\
  $7p_{3/2}$&$  8s_{1/2}$&  11.0522&   11.0181&   10.2979&   10.6244&&   6.2432&    6.2337&    6.0110&    6.0397\\
  $7s_{1/2}$&$  8p_{1/2}$&   0.7965&    0.6773&    0.6704&    0.7222&&   0.2504&    0.0781&    0.1084&    0.1338\\
  $6d_{3/2}$&$  8p_{1/2}$&   3.7239&    3.8052&    3.0878&    2.8125&&   0.3065&    0.4389&    0.0754&    0.0676\\
  $8s_{1/2}$&$  8p_{1/2}$&  12.6289&   12.5513&   11.8450&   11.8417&&   8.5455&    8.4039&    8.0660&    8.1177\\
  $7s_{1/2}$&$  8p_{3/2}$&   1.6259&    1.4908&    1.4225&    1.5019&&   0.8002&    0.5819&    0.6141&    0.6003\\
  $6d_{3/2}$&$  8p_{3/2}$&   1.1810&    1.2095&    0.9565&    0.8218&&   0.0404&    0.0008&    0.1243&    0.1252\\
  $6d_{5/2}$&$  8p_{3/2}$&   3.6698&    3.7574&    3.0607&    2.6305&&   0.0235&    0.1093&    0.2808&    0.2945\\
  $8s_{1/2}$&$  8p_{3/2}$&  16.5110&   16.4367&   15.5457&   15.4156&&  11.4082&   11.2525&   10.7651&   10.8274\\
  $6p_{1/2}$&$  7d_{3/2}$&   1.4465&    1.2125&    1.6900&    1.1299&&   1.1028&    0.6692&    0.5146&    0.5042\\
  $6p_{3/2}$&$  7d_{3/2}$&   0.7804&    0.6653&    0.6691&    0.6110&&   0.5070&    0.3266&    0.2287&    0.2603\\
  $7p_{1/2}$&$  7d_{3/2}$&   4.1553&    4.2092&    4.5174&    4.7613&&   5.1601&    5.0991&    5.4095&    5.4315\\
  $7p_{3/2}$&$  7d_{3/2}$&   2.6765&    2.6800&    2.7508&    2.9263&&   2.8361&    2.7797&    2.9085&    2.9232\\
  $8p_{1/2}$&$  7d_{3/2}$&  24.0271&   23.9277&   22.5109&   22.5898&&  12.5991&   12.4730&   11.4285&   11.4833\\
  $8p_{3/2}$&$  7d_{3/2}$&  10.8862&   10.8523&   10.1532&   10.1562&&   5.4077&    5.3726&    4.8533&    4.8732\\
  $6p_{3/2}$&$  7d_{5/2}$&   2.3233&    1.9804&    1.6868&    1.8107&&   1.5436&    0.9407&    0.7849&    0.7124\\
  $7p_{3/2}$&$  7d_{5/2}$&   7.8353&    7.8510&    7.9661&    8.5204&&   8.2581&    8.1071&    8.4947&    8.5395\\
  $8p_{3/2}$&$  7d_{5/2}$&  32.8077&   32.7038&   30.6899&   30.7015&&  16.5704&   16.4542&   14.9355&   14.9822\\
  $6d_{3/2}$&$  5f_{5/2}$&  15.8106&   15.5875&   13.4045&   13.3048&&   8.8516&    8.2589&    7.3118&    7.3509\\
  $6d_{5/2}$&$  5f_{5/2}$&   4.2672&    4.2077&    3.6468&    3.6153&&   2.4173&    2.2604&    2.0117&    2.0152\\
  $7d_{3/2}$&$  5f_{5/2}$&  24.5718&   24.5696&   25.0077&   24.8311&&  11.3163&   11.3386&   10.9756&   10.9319\\
  $7d_{5/2}$&$  5f_{5/2}$&   6.5453&    6.5450&    6.6575&    6.6158&&   2.9840&    2.9919&    2.8952&    2.8811\\
  $6p_{1/2}$&$  9s_{1/2}$&   0.3651&    0.3228&    0.2837&    0.2987&&   0.2633&    0.2333&    0.1847&    0.1320\\
  $6p_{3/2}$&$  9s_{1/2}$&   0.5325&    0.4372&    0.3874&    0.4143&&   0.4080&    0.3080&    0.2804&    0.2370\\
  $7p_{1/2}$&$  9s_{1/2}$&   1.3385&    1.3441&    1.2659&    1.2721&&   0.8920&    0.9128&    0.8645&    0.8610\\
  $7p_{3/2}$&$  9s_{1/2}$&   1.7190&    1.6982&    1.6157&    1.5851&&   1.2832&    1.2732&    1.2076&    1.2014\\
  $8p_{1/2}$&$  9s_{1/2}$&  12.6067&   12.6187&   12.1123&   12.3331&&   6.4997&    6.5319&    6.3165&    6.3043\\
  $8p_{3/2}$&$  9s_{1/2}$&  20.8624&   20.8557&   19.9495&   20.4187&&  11.1061&   11.1176&   10.8349&   10.8094\\
  $7s_{1/2}$&$  9p_{1/2}$&   0.3474&    0.2690&    0.2763&    0.3061&&   0.0751&    0.0394&    0.0165&    0.0060\\
  $6d_{3/2}$&$  9p_{1/2}$&   1.1097&    1.1597&    0.9717&    0.8978&&   0.0912&    0.1770&    0.0335&    0.0159\\
  $8s_{1/2}$&$  9p_{1/2}$&   1.5895&    1.5400&    1.5143&    1.5865&&   0.4829&    0.3965&    0.4098&    0.4251\\
  $7d_{3/2}$&$  9p_{1/2}$&   7.9981&    8.0423&    6.7565&    6.4442&&   0.9700&    1.0346&    0.4403&    0.4607\\
  $9s_{1/2}$&$  9p_{1/2}$&  20.4396&   20.4054&   19.4813&   19.4532&&  13.2697&   13.1992&   12.7809&   12.7490\\
  $6p_{1/2}$&$  8d_{3/2}$&   0.9468&    0.7747&    0.7275&    0.7000&&   0.6521&    0.3279&    0.1862&    0.1590\\
  $6p_{3/2}$&$  8d_{3/2}$&   0.4911&    0.4073&    0.4608&    0.3664&&   0.2839&    0.1584&    0.0753&    0.1081\\
  $7p_{1/2}$&$  8d_{3/2}$&   2.1894&    2.2102&    2.2981&    2.3602&&   1.8787&    1.8130&    1.7735&    1.7568\\
  $7p_{3/2}$&$  8d_{3/2}$&   1.2490&    1.2442&    1.2445&    1.2819&&   0.8867&    0.8404&    0.7884&    0.7786\\
  $8p_{1/2}$&$  8d_{3/2}$&   5.8636&    5.9043&    6.5642&    6.9811&&   7.9144&    7.9146&    8.5210&    8.5746\\
  $8p_{3/2}$&$  8d_{3/2}$&   3.9589&    3.9679&    4.2072&    4.4888&&   4.4135&    4.3974&    4.6749&    4.6919\\
  $5f_{5/2}$&$  8d_{3/2}$&   2.5635&    2.5573&    4.1317&    4.1826&&   0.2914&    0.2664&    0.5718&    0.5386\\
  $9p_{1/2}$&$  8d_{3/2}$&  39.2901&   39.2403&   37.5983&   37.7089&&  20.5134&   20.4496&   19.1002&   19.0639\\
 \end{tabular}
\end{ruledtabular}
\end{table*}

\begin{table}
\caption{\label{tab-dip-com} Reduced electric-dipole matrix elements
 in Tl~I. SD values $Z^{\text{(SD)}}$ are compared with
theoretical $Z^{\text{(theor)}}$
 and experimental $Z^{\text{(expt)}}$ data
 given in Ref.~\protect\cite{kozlov} and references therein.}
\begin{ruledtabular}
\begin{tabular}{lrrr}
\multicolumn{1}{c}{Transition}&
\multicolumn{1}{c}{$Z^{\text{(SD)}}$ }&
\multicolumn{1}{c}{$Z^{\text{(theor)}}$ }&
\multicolumn{1}{c}{$Z^{\text{(expt)}}$ }\\
\hline
$6p_{1/2}- 7s_{1/2}$& 1.82  &1.77  &1.81$\pm$0.02\\
$6p_{1/2}- 6d_{3/2}$& 2.37  &2.30  &2.30$\pm$0.09\\
$6p_{3/2}- 7s_{1/2}$& 3.39  &3.35  &3.28$\pm$0.04\\
$6p_{3/2}- 6d_{3/2}$& 1.42  &1.40  &1.38$\pm$0.07\\
$6p_{3/2}- 6d_{5/2}$& 4.17  &4.08  &4.0$\pm$0.2\\
$7p_{1/2}- 7s_{1/2}$& 5.90  &5.96  &5.87$\pm$0.08\\
$7p_{1/2}- 6d_{3/2}$& 10.58 &10.86 &              \\
$7p_{3/2}- 7s_{1/2}$& 7.87  &7.98  &7.88$\pm$0.11\\
$7p_{3/2}- 6d_{3/2}$& 4.75  &4.90  &             \\
$7p_{3/2}- 6d_{5/2}$& 14.40 &14.88 &               \\
\end{tabular}
\end{ruledtabular}
\end{table}

\begin{table*}
\caption{\label{tab-osc-tl} Wavelengths $\lambda$ (\AA),
transition rates $A_r$ (s$^{-1}$), oscillator strengths $f$,
and line strengths $S$ (a.u.) for transitions in Tl~I calculated
in the SD approximation. Numbers in brackets represent powers
of 10. }
\begin{ruledtabular}
\begin{tabular}{llrlllllrlll}
\multicolumn{2}{c}{Transition} &
\multicolumn{1}{c}{$\lambda$} &
\multicolumn{1}{c}{$A_r$} &
\multicolumn{1}{c}{$f$} &
\multicolumn{1}{c}{$S$} &
\multicolumn{2}{c}{Transition} &
\multicolumn{1}{c}{$\lambda$} &
\multicolumn{1}{c}{$A_r$} &
\multicolumn{1}{c}{$f$} &
\multicolumn{1}{c}{$S$} \\
\hline
$6p_{1/2}$&$ 7s_{1/2}$&     3808&   6.078[7]&  1.321[-1]& 3.313[ 0]&$7p_{1/2}$&$ 9s_{1/2}$&    11083&   1.204[6]&  2.218[-2]& 1.618[ 0] \\
$6p_{3/2}$&$ 7s_{1/2}$&     5409&   7.375[7]&  1.618[-1]& 1.152[ 1]&$7p_{3/2}$&$ 9s_{1/2}$&    12553&   1.287[6]&  1.520[-2]& 2.513[ 0] \\
$7s_{1/2}$&$ 7p_{1/2}$&    12974&   1.617[7]&  4.080[-1]& 3.485[ 1]&$8p_{1/2}$&$ 9s_{1/2}$&    55279&   9.122[5]&  4.179[-1]& 1.521[ 2] \\
$7s_{1/2}$&$ 7p_{3/2}$&    11409&   2.113[7]&  8.246[-1]& 6.195[ 1]&$8p_{3/2}$&$ 9s_{1/2}$&    69881&   1.238[6]&  4.531[-1]& 4.169[ 2] \\
$6p_{1/2}$&$ 6d_{3/2}$&     2786&   1.320[8]&  3.072[-1]& 5.635[ 0]&$7s_{1/2}$&$ 9p_{1/2}$&     5579&   5.466[5]&  2.550[-3]& 9.368[-2] \\
$6p_{3/2}$&$ 6d_{3/2}$&     3557&   2.268[7]&  4.301[-2]& 2.015[ 0]&$6d_{3/2}$&$ 9p_{1/2}$&    12055&   4.660[5]&  5.077[-3]& 8.060[-1] \\
$7p_{1/2}$&$ 6d_{3/2}$&    52029&   4.024[5]&  3.266[-1]& 1.119[ 2]&$8s_{1/2}$&$ 9p_{1/2}$&    17643&   4.643[5]&  2.167[-2]& 2.517[ 0] \\
$7p_{3/2}$&$ 6d_{3/2}$&   115607&   7.388[3]&  1.480[-2]& 2.253[ 1]&$7d_{3/2}$&$ 9p_{1/2}$&    42808&   5.363[5]&  7.367[-2]& 4.153[ 1] \\
$6p_{3/2}$&$ 6d_{5/2}$&     3546&   1.316[8]&  3.722[-1]& 1.738[ 1]&$9s_{1/2}$&$ 9p_{1/2}$&    83752&   6.526[5]&  6.863[-1]& 3.784[ 2] \\
$7p_{3/2}$&$ 6d_{5/2}$&   105708&   5.930[4]&  1.490[-1]& 2.074[ 2]&$6p_{1/2}$&$ 8d_{3/2}$&     2247&   2.187[7]&  3.311[-2]& 4.900[-1] \\
$6p_{1/2}$&$ 8s_{1/2}$&     2596&   1.634[7]&  1.651[-2]& 2.821[-1]&$6p_{3/2}$&$ 8d_{3/2}$&     2723&   3.367[6]&  3.743[-3]& 1.342[-1] \\
$6p_{3/2}$&$ 8s_{1/2}$&     3253&   1.762[7]&  1.397[-2]& 5.985[-1]&$7p_{1/2}$&$ 8d_{3/2}$&     9500&   3.291[6]&  8.905[-2]& 5.570[ 0] \\
$7p_{1/2}$&$ 8s_{1/2}$&    21983&   3.727[6]&  2.700[-1]& 3.908[ 1]&$7p_{3/2}$&$ 8d_{3/2}$&    10561&   7.066[5]&  1.182[-2]& 1.643[ 0] \\
$7p_{3/2}$&$ 8s_{1/2}$&    28637&   4.869[6]&  2.993[-1]& 1.129[ 2]&$8p_{1/2}$&$ 8d_{3/2}$&    30193&   8.969[5]&  2.452[-1]& 4.874[ 1] \\
$7s_{1/2}$&$ 8p_{1/2}$&     6702&   1.755[6]&  1.182[-2]& 5.215[-1]&$8p_{3/2}$&$ 8d_{3/2}$&    34083&   2.578[5]&  4.489[-2]& 2.015[ 1] \\
$6d_{3/2}$&$ 8p_{1/2}$&    18896&   1.188[6]&  3.179[-2]& 7.910[ 0]&$5f_{5/2}$&$ 8d_{3/2}$&    42017&   1.195[5]&  2.108[-2]& 1.749[ 1] \\
$8s_{1/2}$&$ 8p_{1/2}$&    37523&   2.689[6]&  5.676[-1]& 1.402[ 2]&$9p_{1/2}$&$ 8d_{3/2}$&   323625&   2.125[4]&  6.673[-1]& 1.422[ 3] \\
$7s_{1/2}$&$ 8p_{3/2}$&     6536&   4.092[6]&  5.242[-2]& 2.256[ 0]&$6p_{3/2}$&$ 8d_{5/2}$&     2722&   1.980[7]&  3.298[-2]& 1.182[ 0] \\
$6d_{3/2}$&$ 8p_{3/2}$&    17637&   6.236[4]&  2.908[-3]& 6.754[-1]&$7p_{3/2}$&$ 8d_{5/2}$&    10539&   4.140[6]&  1.034[-1]& 1.435[ 1] \\
$6d_{5/2}$&$ 8p_{3/2}$&    17892&   6.119[5]&  1.958[-2]& 6.919[ 0]&$8p_{3/2}$&$ 8d_{5/2}$&    33852&   1.464[6]&  3.773[-1]& 1.682[ 2] \\
$8s_{1/2}$&$ 8p_{3/2}$&    32862&   3.392[6]&  1.098[ 0]& 2.376[ 2]&$5f_{5/2}$&$ 8d_{5/2}$&    41667&   5.169[3]&  1.345[-3]& 1.107[ 0] \\
$6p_{1/2}$&$ 7d_{3/2}$&     2389&   4.740[7]&  8.115[-2]& 1.277[ 0]&$6d_{3/2}$&$ 6f_{5/2}$&    11455&   5.557[6]&  1.640[-1]& 2.473[ 1] \\
$6p_{3/2}$&$ 7d_{3/2}$&     2935&   7.481[6]&  9.659[-3]& 3.733[-1]&$6d_{5/2}$&$ 6f_{5/2}$&    11562&   3.873[5]&  7.762[-3]& 1.773[ 0] \\
$7p_{1/2}$&$ 7d_{3/2}$&    12689&   5.621[6]&  2.713[-1]& 2.267[ 1]&$7d_{3/2}$&$ 6f_{5/2}$&    36088&   2.250[6]&  6.591[-1]& 3.132[ 2] \\
$7p_{3/2}$&$ 7d_{3/2}$&    14654&   1.378[6]&  4.437[-2]& 8.563[ 0]&$7d_{5/2}$&$ 6f_{5/2}$&    36603&   1.632[5]&  3.278[-2]& 2.370[ 1] \\
$8p_{1/2}$&$ 7d_{3/2}$&   149925&   7.670[4]&  5.169[-1]& 5.103[ 2]&$8d_{3/2}$&$ 6f_{5/2}$&   793651&   1.468[3]&  2.080[-1]& 2.173[ 3] \\
$8p_{3/2}$&$ 7d_{3/2}$&   346021&   1.261[3]&  2.264[-2]& 1.031[ 2]&$8d_{5/2}$&$ 6f_{5/2}$&   943396&   6.218[1]&  8.297[-3]& 1.546[ 2] \\
$6p_{3/2}$&$ 7d_{5/2}$&     2931&   4.396[7]&  8.494[-2]& 3.279[ 0]&$7s_{1/2}$&$ 9p_{3/2}$&     5522&   1.613[6]&  1.475[-2]& 5.363[-1] \\
$7p_{3/2}$&$ 7d_{5/2}$&    14571&   7.925[6]&  3.784[-1]& 7.260[ 1]&$8s_{1/2}$&$ 9p_{3/2}$&    17091&   9.227[5]&  8.082[-2]& 9.094[ 0] \\
$8p_{3/2}$&$ 7d_{5/2}$&   304878&   1.123[4]&  2.348[-1]& 9.426[ 2]&$9s_{1/2}$&$ 9p_{3/2}$&    72622&   8.289[5]&  1.311[ 0]& 6.267[ 2] \\
$6d_{3/2}$&$ 5f_{5/2}$&    16067&   1.441[7]&  8.367[-1]& 1.770[ 2]&$6d_{3/2}$&$ 9p_{3/2}$&    11795&   2.410[4]&  5.028[-4]& 7.809[-2] \\
$6d_{5/2}$&$ 5f_{5/2}$&    16279&   1.023[6]&  4.065[-2]& 1.307[ 1]&$6d_{5/2}$&$ 9p_{3/2}$&    11909&   2.600[5]&  3.685[-3]& 8.668[-1] \\
$7d_{3/2}$&$ 5f_{5/2}$&   377358&   3.875[3]&  1.241[-1]& 6.166[ 2]&$7d_{3/2}$&$ 9p_{3/2}$&    39698&   3.062[4]&  7.235[-3]& 3.782[ 0] \\
$7d_{5/2}$&$ 5f_{5/2}$&   442478&   1.706[2]&  5.008[-3]& 4.377[ 1]&$7d_{5/2}$&$ 9p_{3/2}$&    40323&   2.990[5]&  4.859[-2]& 3.870[ 1] \\
$6p_{1/2}$&$ 9s_{1/2}$&     2326&   7.185[6]&  5.827[-3]& 8.925[-2]&$9p_{3/2}$&$ 8d_{3/2}$&   793651&   2.924[2]&  2.762[-2]& 2.886[ 2] \\
$6p_{3/2}$&$ 9s_{1/2}$&     2839&   7.595[6]&  4.590[-3]& 1.716[-1]&$9p_{3/2}$&$ 8d_{5/2}$&   684932&   2.761[3]&  2.913[-1]& 2.627[ 3] \\
\end{tabular}
\end{ruledtabular}
\end{table*}

\begin{table*}
\caption{\label{tab-osc-pb} Wavelengths $\lambda$ (\AA),
transition rates $A_r$ (s$^{-1}$), oscillator strengths $f$,
and line strengths $S$ (a.u.) for transitions in Pb~II calculated
in the SD approximation. Numbers in brackets represent powers
of 10. }
\begin{ruledtabular}
\begin{tabular}{llrlllllrlll}
\multicolumn{2}{c}{Transition} &
\multicolumn{1}{c}{$\lambda$} &
\multicolumn{1}{c}{$A_r$} &
\multicolumn{1}{c}{$f$} &
\multicolumn{1}{c}{$S$} &
\multicolumn{2}{c}{Transition} &
\multicolumn{1}{c}{$\lambda$} &
\multicolumn{1}{c}{$A_r$} &
\multicolumn{1}{c}{$f$} &
\multicolumn{1}{c}{$S$} \\
\hline
$6p_{1/2}$&$ 7s_{1/2}$&    1696&  2.113[8]&  9.108[-2]& 1.017[ 0]&$7p_{1/2}$&$ 9s_{1/2}$&    3714&  1.466[7]&  3.031[-2]& 7.412[-1]\\
$6p_{3/2}$&$ 7s_{1/2}$&    2231&  3.847[8]&  1.435[-1]& 4.214[ 0]&$7p_{3/2}$&$ 9s_{1/2}$&    4147&  2.050[7]&  2.643[-2]& 1.443[ 0]\\
$6p_{1/2}$&$ 6d_{3/2}$&    1435&  7.318[8]&  4.518[-1]& 4.269[ 0]&$8p_{1/2}$&$ 9s_{1/2}$&   15738&  1.033[7]&  3.835[-1]& 3.974[ 1]\\
$6p_{3/2}$&$ 6d_{3/2}$&    1800&  1.287[8]&  6.255[-2]& 1.483[ 0]&$8p_{3/2}$&$ 9s_{1/2}$&   19242&  1.661[7]&  4.611[-1]& 1.168[ 2]\\
$6p_{3/2}$&$ 6d_{5/2}$&    1783&  7.030[8]&  5.024[-1]& 1.180[ 1]&$6d_{3/2}$&$ 6f_{5/2}$&    3035&  4.477[7]&  9.277[-2]& 3.708[ 0]\\
$7s_{1/2}$&$ 7p_{1/2}$&    6541&  5.647[7]&  3.622[-1]& 1.560[ 1]&$6d_{5/2}$&$ 6f_{5/2}$&    3086&  2.939[6]&  4.196[-3]& 2.558[-1]\\
$6d_{3/2}$&$ 7p_{1/2}$&   21882&  2.727[6]&  9.789[-2]& 2.821[ 1]&$7d_{3/2}$&$ 6f_{5/2}$&   10912&  3.913[7]&  1.048[ 0]& 1.506[ 2]\\
$7s_{1/2}$&$ 7p_{3/2}$&    5524&  8.686[7]&  7.948[-1]& 2.891[ 1]&$7d_{5/2}$&$ 6f_{5/2}$&   11226&  2.754[6]&  5.203[-2]& 1.154[ 1]\\
$6d_{3/2}$&$ 7p_{3/2}$&   13545&  1.040[6]&  2.861[-2]& 5.103[ 0]&$6p_{1/2}$&$ 8d_{3/2}$&     967&  1.415[7]&  3.970[-3]& 2.528[-2]\\
$6d_{5/2}$&$ 7p_{3/2}$&   14620&  7.868[6]&  1.681[-1]& 4.854[ 1]&$6p_{3/2}$&$ 8d_{3/2}$&    1120&  4.206[6]&  7.914[-4]& 1.168[-2]\\
$6p_{1/2}$&$ 8s_{1/2}$&    1125&  9.798[7]&  1.858[-2]& 1.375[-1]&$7p_{1/2}$&$ 8d_{3/2}$&    3432&  3.867[7]&  1.366[-1]& 3.086[ 0]\\
$6p_{3/2}$&$ 8s_{1/2}$&    1337&  1.242[8]&  1.665[-2]& 2.931[-1]&$7p_{3/2}$&$ 8d_{3/2}$&    3799&  5.602[6]&  1.212[-2]& 6.063[-1]\\
$7p_{1/2}$&$ 8s_{1/2}$&    6817&  3.829[7]&  2.668[-1]& 1.197[ 1]&$5f_{5/2}$&$ 8d_{3/2}$&    9016&  2.005[5]&  1.629[-3]& 2.901[-1]\\
$7p_{3/2}$&$ 8s_{1/2}$&    8435&  6.159[7]&  3.284[-1]& 3.648[ 1]&$8p_{1/2}$&$ 8d_{3/2}$&   11671&  2.342[7]&  9.568[-1]& 7.352[ 1]\\
$6d_{3/2}$&$ 5f_{5/2}$&    4421&  2.111[8]&  9.281[-1]& 5.404[ 1]&$8p_{3/2}$&$ 8d_{3/2}$&   13493&  4.539[6]&  1.239[-1]& 2.201[ 1]\\
$6d_{5/2}$&$ 5f_{5/2}$&    4530&  1.475[7]&  4.538[-2]& 4.061[ 0]&$6f_{5/2}$&$ 8d_{3/2}$&  130890&  9.741[4]&  1.668[-1]& 4.312[ 2]\\
$6p_{1/2}$&$ 7d_{3/2}$&    1070&  1.051[8]&  3.608[-2]& 2.542[-1]&$6p_{3/2}$&$ 8d_{5/2}$&    1119&  1.138[7]&  3.203[-3]& 4.719[-2]\\
$6p_{3/2}$&$ 7d_{3/2}$&    1261&  1.714[7]&  4.082[-3]& 6.777[-2]&$7p_{3/2}$&$ 8d_{5/2}$&    3781&  3.553[7]&  1.142[-1]& 5.685[ 0]\\
$7p_{1/2}$&$ 7d_{3/2}$&    5206&  1.059[8]&  8.607[-1]& 2.950[ 1]&$5f_{5/2}$&$ 8d_{5/2}$&    8915&  5.366[3]&  6.393[-5]& 1.126[-2]\\
$7p_{3/2}$&$ 7d_{3/2}$&    6099&  1.908[7]&  1.064[-1]& 8.545[ 0]&$8p_{3/2}$&$ 8d_{5/2}$&   13268&  2.686[7]&  1.063[ 0]& 1.858[ 2]\\
$5f_{5/2}$&$ 7d_{3/2}$&   85985&  9.522[4]&  7.036[-2]& 1.195[ 2]&$6f_{5/2}$&$ 8d_{5/2}$&  112360&  7.147[3]&  1.353[-2]& 3.002[ 1]\\
$6p_{3/2}$&$ 7d_{5/2}$&    1257&  8.638[7]&  3.067[-2]& 5.075[-1]&$7s_{1/2}$&$ 9p_{1/2}$&    2213&  3.402[3]&  2.498[-6]& 3.641[-5]\\
$7p_{3/2}$&$ 7d_{5/2}$&    6005&  1.137[8]&  9.222[-1]& 7.292[ 1]&$6d_{3/2}$&$ 9p_{1/2}$&    2902&  1.047[4]&  6.608[-6]& 2.525[-4]\\
$5f_{5/2}$&$ 7d_{5/2}$&   70472&  8.009[3]&  5.963[-3]& 8.301[ 0]&$8s_{1/2}$&$ 9p_{1/2}$&    6569&  6.460[5]&  4.179[-3]& 1.807[-1]\\
$7s_{1/2}$&$ 8p_{1/2}$&    2789&  8.368[5]&  9.756[-4]& 1.791[-2]&$7d_{3/2}$&$ 9p_{1/2}$&    9361&  2.621[5]&  1.722[-3]& 2.122[-1]\\
$6d_{3/2}$&$ 8p_{1/2}$&    3978&  7.359[4]&  8.728[-5]& 4.572[-3]&$9s_{1/2}$&$ 9p_{1/2}$&   33681&  4.309[6]&  7.329[-1]& 1.625[ 2]\\
$8s_{1/2}$&$ 8p_{1/2}$&   16946&  1.372[7]&  5.906[-1]& 6.590[ 1]&$8d_{3/2}$&$ 9p_{1/2}$&  132450&  1.585[5]&  2.084[-1]& 3.634[ 2]\\
$7d_{3/2}$&$ 8p_{1/2}$&   73529&  3.360[5]&  1.362[-1]& 1.319[ 2]&$7s_{1/2}$&$ 9p_{3/2}$&    2185&  2.647[6]&  3.788[-3]& 5.449[-2]\\
$7s_{1/2}$&$ 8p_{3/2}$&    2702&  9.257[6]&  2.026[-2]& 3.603[-1]&$6d_{3/2}$&$ 9p_{3/2}$&    2853&  8.784[4]&  1.072[-4]& 4.026[-3]\\
$6d_{3/2}$&$ 8p_{3/2}$&    3803&  1.444[5]&  3.131[-4]& 1.568[-2]&$6d_{5/2}$&$ 9p_{3/2}$&    2898&  4.796[5]&  4.025[-4]& 2.304[-2]\\
$6d_{5/2}$&$ 8p_{3/2}$&    3883&  7.502[5]&  1.130[-3]& 8.670[-2]&$8s_{1/2}$&$ 9p_{3/2}$&    6323&  3.575[6]&  4.285[-2]& 1.784[ 0]\\
$8s_{1/2}$&$ 8p_{3/2}$&   14168&  2.088[7]&  1.257[ 0]& 1.172[ 2]&$7d_{3/2}$&$ 9p_{3/2}$&    8869&  1.307[4]&  1.542[-4]& 1.801[-2]\\
$7d_{3/2}$&$ 8p_{3/2}$&   39730&  1.918[5]&  4.539[-2]& 2.375[ 1]&$7d_{5/2}$&$ 9p_{3/2}$&    9075&  2.885[4]&  2.375[-4]& 4.257[-2]\\
$7d_{5/2}$&$ 8p_{3/2}$&   44228&  1.314[6]&  2.569[-1]& 2.245[ 2]&$9s_{1/2}$&$ 9p_{3/2}$&   28082&  6.460[6]&  1.528[ 0]& 2.824[ 2]\\
$6p_{1/2}$&$ 9s_{1/2}$&     988&  1.829[7]&  2.678[-3]& 1.743[-2]&$8d_{3/2}$&$ 9p_{3/2}$&   74239&  8.129[4]&  6.717[-2]& 6.567[ 1]\\
$6p_{3/2}$&$ 9s_{1/2}$&    1149&  3.751[7]&  3.712[-3]& 5.615[-2]&$8d_{5/2}$&$ 9p_{3/2}$&   81900&  5.685[5]&  3.811[-1]& 6.166[ 2]\\
\end{tabular}
\end{ruledtabular}
\end{table*}

\begin{table*}
\caption{\label{tab-ar-com} Transition rates $A_r$
(10$^6$s$^{-1}$) and wavelengths  $\lambda$ (\AA) in Tl~I and
Pb~II. The SD data
                $A_{r}^{\text{(SD)}}$  are compared with
  experimental results $A_{r}^{\text{(expt)}}$
  given in Refs.~\protect\cite{tl-64}--$a$, ~\protect\cite{tl-96}--$b$,
   ~\protect\cite{tl-97}--$c$ for Tl~I and
  Refs.~\protect\cite{pb-01b}--$d$, ~\protect\cite{pb-01a}--$e$ for Pb~II.}
\begin{ruledtabular}
\begin{tabular}{llllllllllll}
\multicolumn{1}{c}{Lower}& \multicolumn{1}{c}{Upper}&
\multicolumn{1}{c}{$\lambda^{\text{(SD)}}$ }&
\multicolumn{1}{c}{$A_{r}^{\text{(SD)}}$ }&
\multicolumn{1}{c}{$\lambda^{\text{(expt)}}$ }&
\multicolumn{1}{c}{$A_{r}^{\text{(expt)}}$ }&
\multicolumn{1}{c}{Lower}& \multicolumn{1}{c}{Upper}&
\multicolumn{1}{c}{$\lambda^{\text{(SD)}}$ }&
\multicolumn{1}{c}{$A_{r}^{\text{(SD)}}$ }&
\multicolumn{1}{c}{$\lambda^{\text{(expt)}}$ }&
\multicolumn{1}{c}{$A_{r}^{\text{(expt)}}$ }\\
\hline \multicolumn{6}{c}{Tl~I}&
 \multicolumn{6}{c}{Pb~II}\\
$6p_{1/2}$&$ 7s_{1/2}$&     3808&     60.8&         & 62.5$\pm$3.1$^a$  &$6p_{3/2}$&$ 7s_{1/2}$&   2231   &385.  &2203.5  &493$\pm$74$^e$   \\
$6p_{3/2}$&$ 7s_{1/2}$&     5409&     73.7&         & 70.5$\pm$3.2$^a$  &$6d_{3/2}$&$ 5f_{5/2}$&   4421   &211.1 &4388.1  &155.7$\pm$15.6$^d$\\
$6p_{1/2}$&$ 6d_{3/2}$&     2786&     132.&         & 126$\pm$10$^a$    &$6d_{5/2}$&$ 5f_{5/2}$&   4530   &14.7  &4243.6  &9.8$\pm$1.0$^d$\\
$6p_{3/2}$&$ 6d_{3/2}$&     3557&     22.7&         & 22.0$\pm$2.3$^a$  &$6d_{3/2}$&$ 8p_1{/2}$&   3978   &0.074 &3947.8  &0.51$\pm$0.06$^d$\\
$6p_{3/2}$&$ 6d_{5/2}$&     3546&     132.&         & 124$\pm$15$^a$    &$6d_{5/2}$&$ 8p_{3/2}$&   3883   &0.750 &3666.5  &0.56$\pm$0.06$^d$\\
$6p_{1/2}$&$ 8s_{1/2}$&     2596&     16.3&         & 17.6$\pm$1.6$^a$  &$6d_{3/2}$&$ 6f_{5/2}$&   3035   &44.8  &3017.4  &40.9$\pm$5.0$^d$\\
$6p_{3/2}$&$ 8s_{1/2}$&     3253&     17.6&         & 17.3$\pm$1.8$^a$  &$6d_{5/2}$&$ 6f_{5/2}$&   3086   &2.94  &2948.4  &3.3$\pm$0.4$^d$ \\
$6p_{1/2}$&$ 7d_{3/2}$&     2389&     47.4&         & 44$\pm$5$^a$      &$7s_{1/2}$&$ 7p_{1/2}$&   6541   &56.5  &6661.8  &57.2$\pm$5.4$^d$\\
$6p_{3/2}$&$ 7d_{3/2}$&     2935&     7.48&         & 7.8$\pm$0.8$^a$   &$7s_{1/2}$&$ 7p_{3/2}$&   5524   &86.9  &5610.0  &84.8$\pm$8.5$^d$\\
$6p_{3/2}$&$ 7d_{5/2}$&     2931&     44.0&         & 42$\pm$5$^a$      &$7s_{1/2}$&$ 8p_{1/2}$&   2789   &0.837 &2806.7  &5.6$\pm$0.6$^d$\\
$6p_{1/2}$&$ 9s_{1/2}$&     2326&     7.18&         & 7.8$\pm$1.0$^a$   &$7s_{1/2}$&$ 8p_{3/2}$&   2702   &9.26  &2718.1  &7.3$\pm$0.8$^d$\\
$6p_{3/2}$&$ 9s_{1/2}$&     2839&     7.59&         & 8.0$\pm$0.8$^a$   &$7p_{1/2}$&$ 8s_{1/2}$&   6817   &38.3  &6793.0  &44.7$\pm$4.5$^d$\\
$6p_{1/2}$&$ 8d_{3/2}$&     2247&     21.9&         & 18.9$\pm$3$^a$    &$7p_{1/2}$&$ 7d_{3/2}$&   5206   &105.9 &5044.1  &90.0$\pm$8.9$^d$\\
$6p_{3/2}$&$ 8d_{3/2}$&     2723&     3.37&         & 3.7$\pm$0.4$^a$   &$7p_{3/2}$&$ 7d_{3/2}$&   6099   &19.1  &5878.2  &13.8$\pm$1.0$^d$\\
$6p_{3/2}$&$ 8d_{5/2}$&     2722&     19.8&         & 17$\pm$2$^a$      &$7p_{3/2}$&$ 7d_{5/2}$&   6005   &113.7 &5545.7  &102.8$\pm$10.6$^d$\\
$7s_{1/2}$&$ 7p_{1/2}$&    12974&     16.2 &13013.2 & 17.1$\pm$0.7$^c$ &$7p_{1/2}$&$ 9s_{1/2}$&   3714   &14.7  &3719.3  &12.0$\pm$2.0$^d$ \\
$7s_{1/2}$&$ 7p_{3/2}$&    11409&     21.1 &11512.8 & 23.7$\pm$0.9$^c$ &$7p_{3/2}$&$ 9s_{1/2}$&   4147   &20.5  &4153.9  &22.3$\pm$2.8$^d$\\
$7s_{1/2}$&$ 8p_{1/2}$&     6702&     1.75 & 6713.8 & 3.7$\pm$0.4$^c$  &$7p_{1/2}$&$ 8d_{3/2}$&   3432   &38.7  &3456.0  &42.1$\pm$6.1$^d$\\
$7s_{1/2}$&$ 8p_{3/2}$&     6536&     4.09 & 6549.9 & 6.2$\pm$0.6$^c$  &$7p_{3/2}$&$ 8d_{3/2}$&   3799   &5.60  &3828.2  &6.3$\pm$0.7$^d$\\
$7s_{1/2}$&$ 9p_{1/2}$&     5579&     0.547& 5564.0 & 0.45$\pm$0.04$^c$&$7p_{3/2}$&$ 8d_{5/2}$&   3781   &35.5  &3715.0  &41.9$\pm$6.2$^d$\\
$7s_{1/2}$&$ 9p_{3/2}$&     5522&     1.61 & 5527.9 & 1.3$\pm$0.1$^c$  &          &           &           &      &        &                \\
$7p_{1/2}$&$ 7d_{3/2}$&    12689&     5.62 &12732.9 & 7.35$\pm$0.62$^b$&          &           &           &      &        &                \\
$7p_{3/2}$&$ 7d_{3/2}$&    14654&     1.38 &14593.9 & 1.74$\pm$0.15$^b$&          &           &           &      &        &                \\
$7p_{1/2}$&$ 9s_{1/2}$&    11083&     1.20 &11100.3 & 1.15$\pm$0.11$^b$&          &           &           &      &        &                \\
$7p_{3/2}$&$ 9s_{1/2}$&    12553&     1.29 &12488.7 & 1.44$\pm$0.14$^b$&          &           &           &      &        &                \\
$7p_{1/2}$&$ 8d_{3/2}$&     9500&     3.29 & 9509.4 & 3.25$\pm$0.23$^b$&          &           &           &      &        &                \\
$7p_{3/2}$&$ 8d_{3/2}$&    10561&     0.707& 10510.7& 1.19$\pm$0.09$^b$&          &           &           &      &        &                \\
\end{tabular}
\end{ruledtabular}
\end{table*}

\begin{table}
\caption{\label{tab-life} Lifetimes ${\tau}$
 of the $nl$  levels
 in Tl~I and Pb~II in ns.
 The SD data are compared with experimental
results for Tl~I from Ref.~\protect\cite{tl-71} ($7s_{1/2}$),
Ref.~\protect\cite{tl-6p86}  and for Pb~II from
Ref.~\protect\cite{pb-01b} and references therein.}
\begin{ruledtabular}
\begin{tabular}{llllll}
\multicolumn{1}{c}{Level}                   &
\multicolumn{1}{c}{$\tau^{(\rm SD)}$}            &
\multicolumn{1}{c}{$\tau^{\text {expt}}$}   &
\multicolumn{1}{c}{Level}                   &
\multicolumn{1}{c}{$\tau^{(\rm SD)}$}            &
\multicolumn{1}{c}{$\tau^{\text {expt}}$}   \\
\hline \multicolumn{3}{c}{Tl~I, $Z$=81}&
\multicolumn{3}{c}{Pb~II, $Z$=82}\\
$ 7s_{1/2}$   &7.43    &7.45$\pm$0.2   &$ 7s_{1/2}$   &1.68    &7.2$\pm$0.9 \\
$ 7p_{1/2}$   &61.8    &63.1$\pm$1.7   &$ 7p_{1/2}$   &16.9   &15.2$\pm$1.7 \\
$ 7p_{3/2}$   &47.3    &48.6$\pm$1.3   &$ 7p_{3/2}$   &10.4   &10.3$\pm$1.2 \\
$ 8p_{1/2}$   &177.6   &184.1$\pm$4.4  &$ 7d_{3/2}$   &4.05   &3.4$\pm$0.4 \\
$ 8p_{3/2}$   &123.5   &127.7$\pm$4.9  &$ 5f_{5/2}$   &4.43   &5.9$\pm$0.6 \\
$ 9p_{1/2}$   &375.1   &391.1$\pm$21.8 &$ 6f_{5/2}$   &11.5   &11.6$\pm$1.5 \\
$ 9p_{3/2}$   &251.3   &273.6$\pm$13.5 &              &       &             \\
\end{tabular}
\end{ruledtabular}
\end{table}

\begin{table*}
\caption{\label{tab-dip-e2m1} Reduced matrix elements of the
electric-quadrupole and magnetic-dipole operator in first, second,
third, and all orders of perturbation theory in  Tl~I and Pb~II.}
\begin{ruledtabular}
\begin{tabular}{llrrrrrrrr}
\multicolumn{2}{c}{Transition}&
\multicolumn{1}{c}{$Z^{(1)}$ }&
\multicolumn{1}{c}{$Z^{(2)}$ }&
\multicolumn{1}{c}{$Z^{(3)}$ }&
\multicolumn{1}{c}{$Z^\text{{(SD)}}$ }&
\multicolumn{1}{c}{$Z^{(1)}$ }&
\multicolumn{1}{c}{$Z^{(2)}$ }&
\multicolumn{1}{c}{$Z^{(3)}$ }&
\multicolumn{1}{c}{$Z^\text{{(SD)}}$ }\\
\hline
\multicolumn{5}{c}{Tl~I}&
\multicolumn{4}{c}{Pb~II}\\
\multicolumn{9}{c}{ Electric-quadrupole transitions}\\
$6p_{1/2}$&$6p_{3/2}$&  15.2950& 15.4716&  12.4935&  13.1672&  9.1374&  9.1936&  8.1941&  8.2784\\
$6p_{1/2}$&$7p_{3/2}$&   7.1787&  7.2776&   5.7125&   6.3019&  4.0244&  4.1182&  3.8682&  4.0294\\
$6p_{3/2}$&$7p_{1/2}$&  15.8767& 15.9107&  11.7256&  13.2822&  8.0165&  8.0706&  7.3327&  7.6247\\
$7p_{1/2}$&$7p_{3/2}$& 127.7551&128.1446& 110.4106& 114.0819& 53.1141& 53.1487& 47.8697& 48.6564\\
\multicolumn{9}{c}{ Magnetic-dipole transitions}\\
$6p_{1/2}$&$6p_{3/2}$&   1.1354&  1.1358&   1.0959&   1.1366&  1.1372&  1.1375&  1.0963&  1.1371\\
$6p_{1/2}$&$7p_{3/2}$&   0.1022&  0.1016&   0.0771&   0.1020&  0.0971&  0.0965&  0.0832&  0.0971\\
$6p_{3/2}$&$7p_{1/2}$&   0.1228&  0.1218&   0.1137&   0.1077&  0.1169&  0.1160&  0.1197&  0.1161\\
$7p_{1/2}$&$7p_{3/2}$&   1.1387&  1.1380&   1.1345&   1.1384&  1.1391&  1.1393&  1.1261&  1.1379\\
\end{tabular}
\end{ruledtabular}
\end{table*}

\begin{table}
\caption{\label{tab-ar-e2m1} Wavelengths $\lambda$ (\AA) and
transition rates for electric-quadrupole $A^{E2}_{r}$ and
magnetic-dipole $A^{M1}_{r}$
 (s$^{-1}$) transitions in Tl~I and Pb~II calculated in
the SD approximation.
The SD data ($a$) are compared with theoretical calculations
  given in Refs.~\protect\cite{tl-forb-77}--($b$) 
and  ~\protect\cite{forb-96}--($c$).
Numbers in brackets represent
powers of 10. }
\begin{ruledtabular}
\begin{tabular}{lllrll}
\multicolumn{2}{c}{Transition} &
\multicolumn{1}{c}{} &
\multicolumn{1}{c}{$\lambda$} &
\multicolumn{1}{c}{$A^{E2}_{r}$} &
\multicolumn{1}{c}{$A^{M1}_{r}$} \\
\hline
\multicolumn{6}{c}{Tl~I}\\
$6p_{1/2}$&$6p_{3/2}$& $a$& 12862&  0.1379&   4.094\\
          &          & $b$&      &  0.158&    4.085\\
          &          & $c$&      &  0.1978&   4.268\\
$6p_{1/2}$&$7p_{3/2}$& $a$&  2861&  58.05&    2.996\\
          &          & $b$&      &  55.2&     3.31\\
$6p_{3/2}$&$7p_{1/2}$& $a$&  3819&  121.5&    2.810\\
          &          & $b$&      &  72.8&     2.18]\\
$7p_{1/2}$&$7p_{3/2}$& $a$& 99900&  3.66[-4]&   8.765[-3]\\
          &          & $b$&      &  3.69[-4]&   8.706[-3]\\
\multicolumn{6}{c}{Pb~II}\\
$6p_{1/2}$&$6p_{3/2}$& $a$&    7074& 1.083 &  24.63\\
         &           & $c$&        & 1.365 &  25.2\\
$6p_{1/2}$&$7p_{3/2}$& $a$&    1298& 1236 &  29.13\\
$6p_{3/2}$&$7p_{1/2}$& $a$&    1663& 2557 &  395.3\\
$7p_{1/2}$&$7p_{3/2}$& $a$&   35549& 0.0117 &  0.1944\\
\end{tabular}
\end{ruledtabular}
\end{table}

\begin{table}
\caption{\label{contr-hyp} Contributions to the SD values of the
$^{205}$Tl hyperfine constants in MHz. Designations are from
Refs.~\protect\cite{blundell-li,Safronova}.}
\begin{ruledtabular}
\begin{tabular}{lrrrrr}
\multicolumn{1}{c}{Terms}& \multicolumn{1}{c}{$6p_{1/2}$ }&
\multicolumn{1}{c}{$6p_{3/2}$ }& \multicolumn{1}{c}{$6d_{3/2}$ }&
\multicolumn{1}{c}{$6d_{5/2}$ }& \multicolumn{1}{c}{$7s_{1/2}$
}\\\hline
$A^{(\rm DHF)}$&   17413.9&    1302.0&     20.71&     8.660&    7380.9\\[0.4pc]
$Z^{(a)}$&         2217.8&   -1540.7&     33.05&   -21.613&    2423.2\\
$Z^{(b)}$&         -690.2&     -44.6&      1.46&     0.597&   -1919.1\\
$Z^{(c)}$&         3361.3&     299.1&     22.93&     9.234&    4247.4\\
$Z^{(d)}$&          162.4&      17.2&      6.53&     2.531&     611.0\\
$Z^{(e)}$&            6.9&       0.4&      0.03&     0.011&     124.8\\
$Z^{(f)}$&          -35.4&      -2.5&      0.63&     0.243&    -298.1\\
$Z^{(g)}$&           -8.4&      -0.5&      0.00&     0.001&    -116.6\\
$Z^{(h)}$&           93.1&     -29.7&     12.59&    -8.563&     436.0\\
$Z^{(i)}$&           83.9&    -108.2&      1.45&    -1.337&     120.2\\
$Z^{(j)}$&          -52.3&      90.9&     -2.93&     2.996&    -128.5\\
$Z^{(k)}$&          -19.1&       0.0&      0.08&    -0.046&    -115.3\\
$Z^{(l)}$&           -5.7&       0.2&      0.07&     0.033&       1.6\\
$Z^{(m)}$&         -299.6&     -21.7&     -0.32&    -0.127&     -18.0\\
$Z^{(n)}$&         -347.0&     269.3&   -173.53&   195.598&     245.9\\
$Z^{(o)}$&           27.5&      89.8&      0.43&     0.339&      42.3\\
$Z^{(p)}$&          719.3&     190.5&     30.11&    35.359&     110.5\\
$Z^{(q)}$&          357.6&      -0.3&      0.02&     0.069&      10.3\\
$Z^{(r)}$&           92.8&     -57.9&      1.99&    -1.695&     109.9\\
$Z^{(s)}$&         -292.6&     -22.6&      1.51&     0.624&    -294.9\\
$Z^{(t)}$&        -1222.8&     -72.0&     -0.34&    -0.148&    -112.1\\
$Z^{\rm (den)}$&   -174.0&      -5.7&      1.48&    -7.343&    -265.0\\[0.4pc]
$A^{\rm (SD)}$&  21389.6&     353.3&    -42.06&   215.423&   12596.5\\
\end{tabular}
\end{ruledtabular}
\end{table}
\begin{table}
\caption{\label{tab-hyp} Hyperfine constants $A$ (in
MHz) for the $np_j$ with $n=6-9$,   $ns_{1/2}$ with $n=7-9$, and
$6d_j$ levels in
 $^{205}$Tl ($I$=1/2, $\mu$=1.6382135).
 The SD data are compared with theoretical and experimental
results from Ref.~\protect\cite{kozlov} - ($a$),
 Ref.~\protect\cite{tl-forb-77} - ($b$),  Ref.~\protect\cite{tl-pra88} -
 ($c$),
 and Ref.~\protect\cite{tl-hypd} - ($d$).}
\begin{ruledtabular}
\begin{tabular}{rrrrr}
\multicolumn{1}{c}{Level} & \multicolumn{1}{c}{$A^{(\rm DHF)}$} &
\multicolumn{1}{c}{$A^{(\rm SD)}$} & \multicolumn{1}{c}{$A^{(\rm
theor)}$} &
\multicolumn{1}{c}{$A^{(\rm expt)}$} \\
\hline
$6p_{1/2}$ &  17414    & 21390 & 21663$^a$  &  21310.8$\pm$0.0$^c$\\
$6p_{3/2}$ &  1302     & 353   &   248$^a$  &    265.0$\pm$0.0$^c$\\
$7s_{1/2}$ &  7381     & 12596 & 12666$^a$  &  12297.2$\pm$1.6$^d$\\
$7p_{1/2}$ &  1942     & 2248  &  2193$^a$  &   2155.5$\pm$0.6$^c$\\
$7p_{3/2}$ &  187.9    & 294.3 &   295$^a$  &    311.4$\pm$0.3$^c$\\
$6d_{3/2}$ &  20.7     &-42.1  &   -41$^a$  &    -42.9$\pm$0.4$^d$\\
$6d_{5/2}$ &  8.66     & 215.4 &   183$^a$  &    226.9$\pm$0.3$^d$\\
$8s_{1/2}$ &  2479     & 3908  &  4320$^b$  & 3870.8$\pm$1.4$^d$\\
$8p_{1/2}$ &  730      & 836   &   706$^c$  &  788.5$\pm$0.9$^c$\\
$8p_{3/2}$ &  72.1     & 122   &    61$^c$  &  130.2$\pm$0.5$^c$\\
$9s_{1/2}$ &  1127     & 1657  &  1900$^b$  &  1779.4$\pm$1.2$^d$\\
$9p_{1/2}$ &  356      & 401   &            &  378.4$\pm$0.8$^c$\\
$9p_{3/2}$ &  35.7     & 62.1  &            &   67.1$\pm$0.2$^c$\\
$7d_{3/2}$ &  11.0     &-45.6  &            &    -56.3$\pm$0.6$^d$\\
$7d_{5/2}$ &  4.55    & 135    &            &   180.2$\pm$0.2$^d$\\
$8d_{3/2}$ &  6.10    &-32.0   &            &    -42.3$\pm$0.4$^d$\\
$8d_{5/2}$ &  2.53    & 83.1   &            &   130.6$\pm$0.2$^d$\\
\end{tabular}
\end{ruledtabular}
\end{table}


\end{document}